\RequirePackage{fix-cm}
\documentclass[aps,prl,twocolumn,nofootinbib,longbibliography,amsfonts,amssymb,amsmath,superscriptaddress,nobibnotes]{revtex4-2}
\usepackage[T1]{fontenc}
\usepackage{lmodern} %
\rmfamily
\DeclareFontShape{T1}{lmr}{b}{sc}{<->ssub*cmr/bx/sc}{}
\DeclareFontShape{T1}{lmr}{bx}{sc}{<->ssub*cmr/bx/sc}{}
\usepackage[dvipsnames,table]{xcolor}
\usepackage{microtype}
\usepackage{IEEEtrantools}
\usepackage{mathrsfs}
\usepackage{amsthm}
\usepackage{enumitem}
\usepackage{varwidth}
\usepackage[normalem]{ulem}
\usepackage{graphicx}
\usepackage{stackengine}
\usepackage{cprotect}
\usepackage[caption=false]{subfig}
\usepackage{adjustbox}
\usepackage{multirow}
\usepackage[pdftex]{hyperref}
\usepackage[USenglish]{babel}
\usepackage{microtype}
\usepackage{layouts}
\usepackage{algorithm}
\usepackage{algpseudocode}
\usepackage{varwidth}
\usepackage{xspace}

\usepackage{tikz}
\usetikzlibrary{shapes,arrows.meta,positioning,graphs,fit,backgrounds,calc}

\newcommand{\Mc}{\mathcal{M}}  %
\newcommand{\Mcp}{{\widetilde \Mc}}  %

\newcommand{\supp}{Supplementary Material\xspace}
\newcommand{\citesearch}{\cite{Finn:1992wt,Nitz:2018rgo,Cannon:2020qnf,Adams:2015ulm,Chu:2020pjv,Sachdev:2020lfd, Nitz:2020vym,Kovalam:2021bgg}\xspace}
\newcommand{\citedingo}{\cite{Green:2020hst,Green:2020dnx,Dax:2021tsq,Dax:2022pxd}\xspace}
\newcommand{\citeheterodyning}{\cite{Cornish:2010kf,Cornish:2021lje,Zackay:2018qdy}\xspace}
\newcommand{\citemultibanding}{\cite{Vinciguerra:2017ngf,Morisaki:2021ngj}\xspace}

\definecolor{tabblue}{RGB}{31, 119, 180}
\definecolor{taborange}{RGB}{255, 127, 14}
\definecolor{tabgreen}{RGB}{44, 160, 44}
\definecolor{tabred}{RGB}{214, 39, 40}
\definecolor{tabpurple}{RGB}{148, 103, 189}
\definecolor{tabbrown}{RGB}{140, 86, 75}
\definecolor{tabpink}{RGB}{227, 119, 194}
\definecolor{tabgray}{RGB}{127, 127, 127}
\definecolor{tabolive}{RGB}{188, 189, 34}
\definecolor{tabcyan}{RGB}{23, 190, 207}
\definecolor{teal}{RGB}{0, 128, 128}

\renewcommand{\sec}[1]{\textbf{\emph{#1.---}}}

\begin{document}

\title{Real-time gravitational-wave inference for binary neutron stars using machine learning}

\author{Maximilian Dax}
\email{maximilian.dax@tuebingen.mpg.de}
\affiliation{Max Planck Institute for Intelligent Systems, Max-Planck-Ring 4, 72076 T\"ubingen, Germany}
\author{Stephen R. Green}
\email{stephen.green2@nottingham.ac.uk}
\affiliation{School of Mathematical Sciences, University of Nottingham, University Park, Nottingham NG7 2RD, United Kingdom}
\author{Jonathan Gair}
\affiliation{Max Planck Institute for Gravitational Physics (Albert Einstein Institute), Am M\"uhlenberg 1, 14476 Potsdam, Germany}
\author{Nihar Gupte}
\affiliation{Max Planck Institute for Gravitational Physics (Albert Einstein Institute), Am M\"uhlenberg 1, 14476 Potsdam, Germany}
\affiliation{Department of Physics, University of Maryland, College Park, MD 20742, USA}
\author{Michael P\"urrer}
\affiliation{Department of Physics, East Hall, University of Rhode Island, Kingston, RI 02881, USA}
\affiliation{Center for Computational Research, Carothers Library, University of Rhode Island, Kingston, RI 02881, USA}
\author{Vivien Raymond}
\affiliation{Gravity Exploration Institute, Cardiff University, Cardiff CF24 3AA, United Kingdom}
\author{Jonas Wildberger}
\affiliation{ELLIS Institute T\"ubingen, Maria-von-Linden-Straße 2,
72076 T\"ubingen, Germany}
\author{Jakob H.~Macke}
\affiliation{Max Planck Institute for Intelligent Systems,  Max-Planck-Ring 4, 72076 T\"ubingen, Germany}
\affiliation{Machine Learning in Science, University of T\"ubingen \& Tübingen AI Center, 72076 T\"ubingen, Germany}
\author{Alessandra Buonanno}
\affiliation{Max Planck Institute for Gravitational Physics (Albert Einstein Institute), Am M\"uhlenberg 1, 14476 Potsdam, Germany}
\affiliation{Department of Physics, University of Maryland, College Park, MD 20742, USA}
\author{Bernhard Sch\"{o}lkopf}
\affiliation{Max Planck Institute for Intelligent Systems,  Max-Planck-Ring 4, 72076  T\"ubingen, Germany}
\affiliation{ELLIS Institute T\"ubingen, Maria-von-Linden-Straße 2,
72076 T\"ubingen, Germany}

\begin{abstract}
    Mergers of binary neutron stars (BNSs) emit signals in both the gravitational-wave (GW) and electromagnetic (EM) spectra. Famously, the 2017 multi-messenger observation of GW170817~\cite{TheLIGOScientific:2017qsa,LIGOScientific:2017ync} led to scientific discoveries across cosmology~\cite{LIGOScientific:2017adf}, nuclear physics~\cite{LIGOScientific:2017zic,Abbott:2018wiz,Abbott:2018exr}, and gravity~\cite{LIGOScientific:2018dkp}. Central to these results were the sky localization and distance obtained from GW data, which, in the case of GW170817, helped to identify the associated EM transient, AT 2017gfo~\cite{Coulter:2017wya}, 11 hours after the GW signal. Fast analysis of GW data is critical for directing time-sensitive EM observations; however, due to challenges arising from the length and complexity of signals, it is often necessary to make approximations that sacrifice accuracy. Here, we present a machine learning framework that performs complete BNS inference in just one second without making any such approximations. Our approach enhances multi-messenger observations by providing (i) accurate localization even before the merger; (ii) improved localization precision by $\sim30\%$ compared to approximate low-latency methods; and (iii) detailed information on luminosity distance, inclination, and masses, which can be used to prioritize expensive telescope time. Additionally, the flexibility and reduced cost of our method open new opportunities for equation-of-state studies. Finally, we demonstrate that our method scales to extremely long signals, up to an hour in length, thus serving as a blueprint for data analysis for next-generation ground- and space-based detectors.
\end{abstract}

\maketitle 

\sec{Introduction}Fast and accurate inference of binary neutron stars (BNSs) from gravitational-wave (GW) data is a critical challenge facing multi-messenger astronomy. For a BNS, the GW signal is visible by the LIGO-Virgo-KAGRA (LVK)~\cite{LIGOScientific:2014pky,VIRGO:2014yos,Aso:2013eba} observatories minutes before any electromagnetic counterpart, and encodes information on source characterization, distance, sky location, and orientation necessary for pointing and prioritizing optical telescopes. However, the length of BNS signals makes conventional Bayesian inference techniques~\cite{Veitch:2014wba,Ashton:2018jfp} too slow to be useful in low-latency applications.  Instead, once a GW signal is identified by detection pipelines~\citesearch, approximate algorithms are used for providing initial alerts (e.g., Bayestar~\cite{Singer:2015ema}, which uses the signal-to-noise [SNR] time series rather than the complete strain data and gives localization in seconds). Other methods focus on accelerating likelihood evaluations without incurring loss of precision (e.g., using reduced-order quadratures), with the state-of-the-art delivering localization in six minutes, and full inference in two hours~\cite{Morisaki:2023kuq}.

Simulation-based machine learning offers a powerful alternative for GW inference (see \supp for related work). With simulation-based inference (SBI)~\cite{cranmer2020frontier}, neural networks are trained to encode probabilistic estimates of astrophysical source parameters conditional on data. Trained networks then enable extremely fast analysis for new data sets, amortizing upfront training costs across observations. In past work, we developed the \textsc{Dingo} framework for binary black holes (BBHs)~\citedingo, which performs accurate inference in seconds, including strong accuracy guarantees when coupled with importance sampling. However, when applied to BNS, machine learning approaches such as \textsc{Dingo} are beset by the same challenges facing traditional methods due to long signal durations. Indeed, \textsc{Dingo} becomes unreliable even for low-mass BBHs (chirp masses $\lesssim 15~\text{M}_\odot$), with signals longer than roughly 16~s. A BNS lasts for hundreds of seconds for the LVK and will reach hours for next-generation detectors (XG, e.g., Cosmic Explorer~\cite{Reitze:2019iox} and Einstein Telescope~\cite{Punturo:2010zza}). From the neural network perspective, this corresponds to time or frequency series input with up to tens of millions of dimensions, a thousand-fold increase over BBH. 

\tikzstyle{decision} = [diamond, draw, fill=blue!20, 
    text width=4.5em, text badly centered, node distance=3cm, inner sep=0pt]
\tikzstyle{block} = [rectangle, draw, fill=blue!20, 
    minimum width=1cm, align=center, minimum height=4em]
\tikzstyle{cloud} = [rectangle, draw, rounded corners,fill=red!20, align=center, 
minimum height=2em]

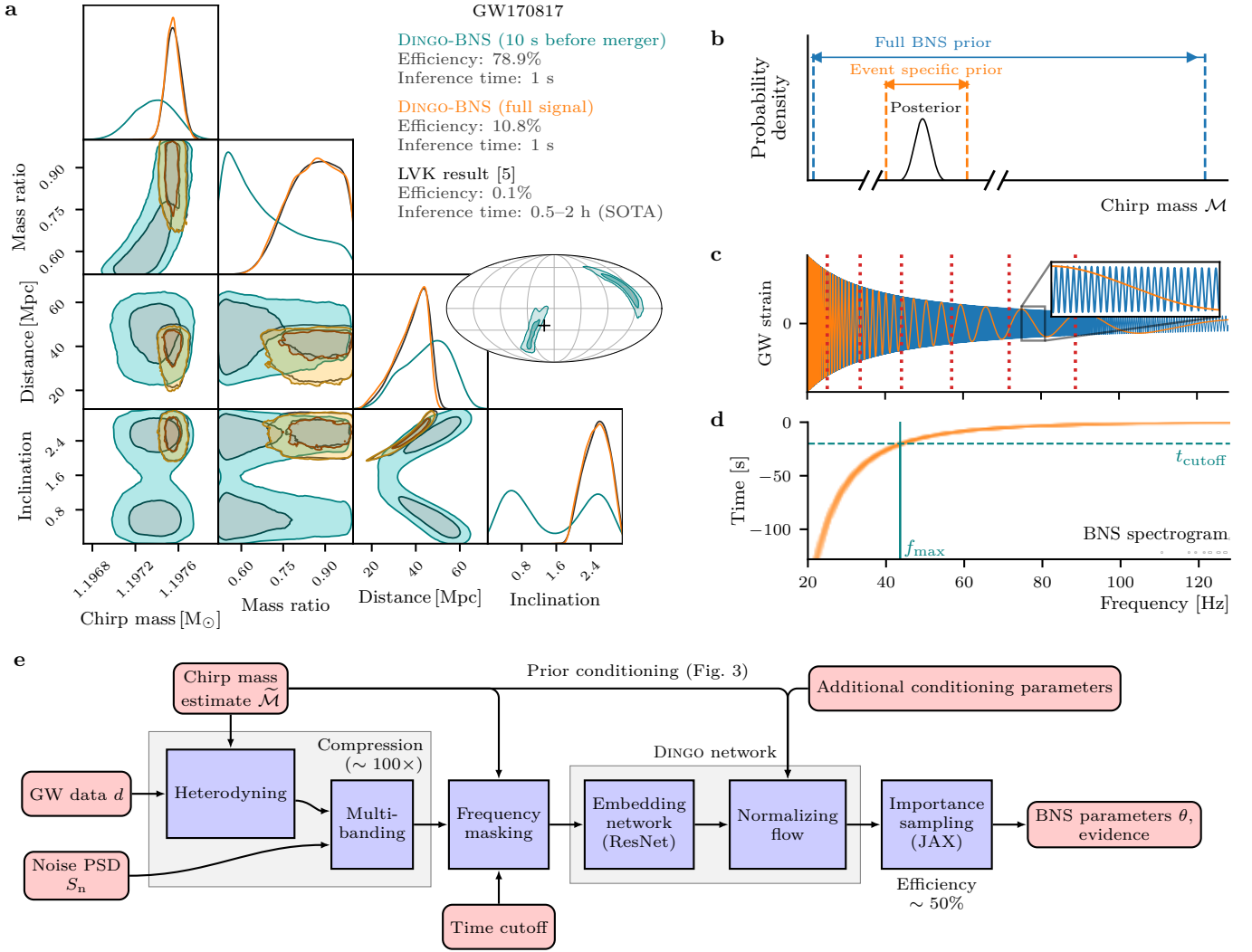
\begin{figure*}\vspace{-1.5ex}
  \begin{minipage}[t]{.5\textwidth}
  \subfloat{
    \stackinset{l}{0cm}{t}{0cm}{\textbf{a}}{
    \stackinset{r}{-.5cm}{t}{0mm}{
      \scriptsize
      \begin{varwidth}{\linewidth}
      GW170817
      \begin{itemize}[label={}, leftmargin=*]
        \item \textcolor{teal}{\textsc{Dingo-BNS} (10 s before merger)}\\\textcolor{darkgray}{Efficiency: 78.9\%\\Inference time: 1 s}
        \item \textcolor{taborange}{\textsc{Dingo-BNS} (full signal)}\\\textcolor{darkgray}{Efficiency: 10.8\%\\Inference time: 1 s}
        \item LVK result \cite{Abbott:2018wiz}\\\textcolor{darkgray}{Efficiency: 0.1\%\\Inference time: 0.5--2 h (SOTA)}
      \end{itemize}
      \end{varwidth}
    }
    {\stackinset{r}{-.5cm}{b}{3.9cm}{\includegraphics[width=0.36\textwidth]{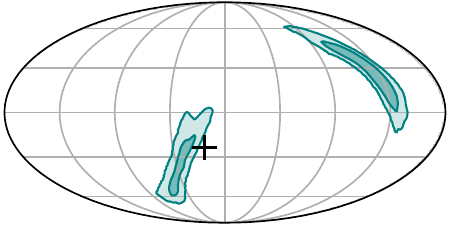}}
    {\raggedright
      \includegraphics[width=\textwidth]{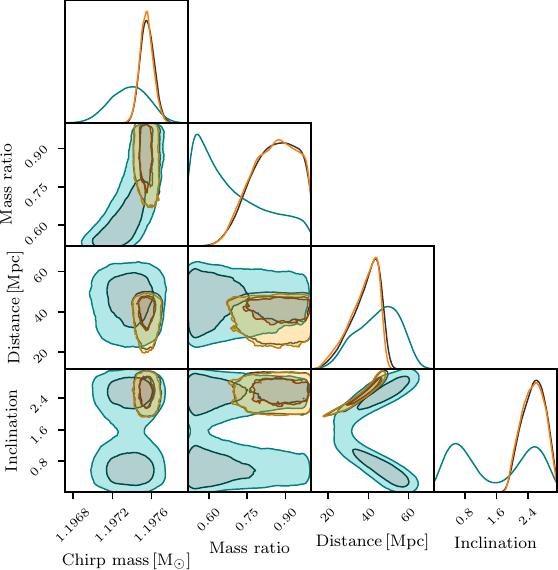}
    }}}
  }
  \end{minipage}
  \hfill
  \begin{minipage}[t]{.422\textwidth}
    \subfloat[\label{fig:prior-conditioning-high-level}]{
      \stackinset{l}{0cm}{t}{0cm}{\textbf{b}}{\raggedright\hspace{.055\textwidth}
        \includegraphics[width=.945\textwidth]{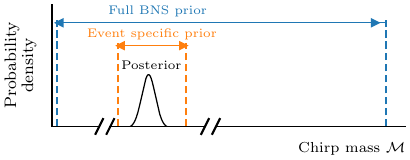}}}\\\vspace{0.05cm}
    \subfloat[\label{fig:heterodyned-multibanding}]{
      \stackinset{l}{0cm}{t}{0cm}{\textbf{c}}{\raggedright\hspace{.07\textwidth}
        \includegraphics[width=.93\textwidth]{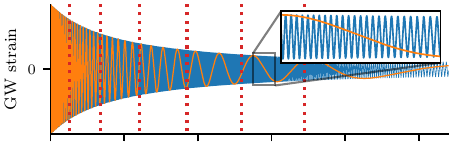}}}\\\vspace{-0.3cm}
    \subfloat[\label{fig:fmask}]{
      \stackinset{l}{0cm}{t}{0cm}{\textbf{d}}{
        \stackinset{r}{0.1cm}{t}{1.7cm}{\scalebox{0.85}{BNS spectrogram}}{
        \raggedright \includegraphics[width=.99\textwidth]{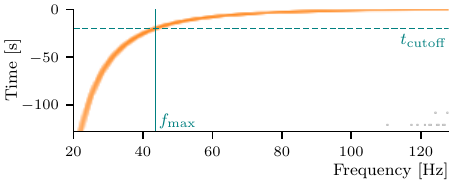}}}}\\
  \end{minipage}\\
    \subfloat[\label{fig:flowchart}]{
       \stackinset{l}{0cm}{t}{0cm}{\textbf{e}}{
    \begin{tikzpicture}[node distance = .5cm, thick, >={Latex[scale=.6]}, font=\scriptsize]
      \node [cloud] (data) {GW data $d$};
      \node [cloud, below=of data] (psd) {Noise PSD\\$S_\text{n}$};
      \node [block, right=of data] (heterodyning) {Heterodyning};
      \node [inner sep=0, fit=(data) (heterodyning) (psd)] (ingroup) {};
      \node [cloud, above=of heterodyning] (chirp) {Chirp mass\\estimate $\Mcp$};
      \node [block, right=of ingroup] (multibanding) {Multi-\\banding};
      \node [block, right=of multibanding] (fmask) {Frequency\\masking};
      \node [block, right=of fmask] (embedding) {Embedding\\network\\(ResNet)};
      \node [block, right=of embedding] (flow) {Normalizing\\flow};
      \node [block, right=of flow, label={[align=center]south:{Efficiency\\$\sim 50\%$}}] (is) {Importance\\sampling\\(JAX)};
      \node [cloud, right=of is] (weighted) {BNS parameters $\theta$,\\evidence};
      \node [cloud, below=of fmask] (tcutoff) {Time cutoff};
      \begin{scope}[on background layer]
        \node [draw, thin, gray, fill=gray!10, inner sep=2.5mm, fit=(heterodyning) (multibanding), label={[anchor=north east, align=right]north east:{Compression\\($\sim 100\times$)}}] (compression) {};
        \node [draw, thin, gray, fill=gray!10, inner sep=2mm, fit=(embedding) (flow), label={\textsc{Dingo} network}] (nn) {};
      \end{scope}
      \graph{
          (chirp) -> (heterodyning);
          (data) -> (heterodyning);
          (multibanding) -> (fmask) -> (embedding) -> (flow) -> (is) -> (weighted);
          (tcutoff) -> (fmask);
        };
      \draw [->, rounded corners] (chirp) -| (flow) coordinate[pos=0.5] (bend) node[pos=0.35, above] {Prior conditioning (Fig.~\ref{fig:prior-conditioning})};
      \node [cloud, right=of bend, xshift=-2mm] (additional) {Additional conditioning parameters};
      \draw [->, rounded corners] (additional) -| (flow);
      \draw [->, rounded corners] (chirp) -| (fmask);
      \draw [->] ([yshift=-1mm]heterodyning.east) to[out=0,in=180] ([yshift=2mm]multibanding.west);
      \draw [->] (psd) to[out=0,in=180] ([yshift=-3mm]multibanding.west);
    \end{tikzpicture}}
  }
\caption{\label{fig:dingo-bns} 
    \textbf{Real-time GW inference for BNS is enabled by several innovations.}
    (a) \textsc{Dingo-BNS} estimates all BNS parameters in just one second (orange), reproducing LVK results~\cite{Abbott:2018wiz} (black) three orders of magnitude faster than existing methods~\cite{Morisaki:2023kuq,Wong:2023lgb}. \textsc{Dingo-BNS} can also analyze partial data before the merger occurs (teal). 
    Fast analysis results are crucial for directing  electromagnetic searches for prompt or even precursor signals. 
    Note that GW170817 overlapped with a loud glitch, which could explain why the true sky position lies in the tail of the pre-merger distribution.
    (b) For a given event, the chirp mass posterior (black) is tightly constrained compared to the prior (blue), so a restricted chirp mass prior (orange) is sufficient, and moreover simplifies analysis. With our prior-conditioning technique, we train a single neural network that can be instantly tuned to an event-specific prior lying anywhere within the full volume.
    (c) We compress data by a factor of $\sim 100$ by first factoring out (``heterodyning'') the predominant phase evolution of the signal (blue), based on a chirp mass estimate $\Mcp$ associated to the event-specific prior. The resulting simplified signal (orange) is down-sampled in resolution, reducing data dimensionality (coarser resolution at high frequencies; bands indicated by dotted red lines). 
    (d) To enable pre-merger inference, we mask out the strain frequency series according to the cut-off time. 
    (e) All of these components are integrated into a single neural network that can be trained end-to-end and produce $10^5$ weighted samples per second, with typical sampling efficiencies of 50\%.
    } 
\end{figure*}

\begin{figure*}\vspace{-1.5ex}
    \begin{minipage}[t]{.49\textwidth}
    \subfloat[\label{fig:GW170817-injection}]{\stackinset{l}{0cm}{t}{0cm}{\textbf{a}}{\raggedright
    \includegraphics[width=\textwidth]{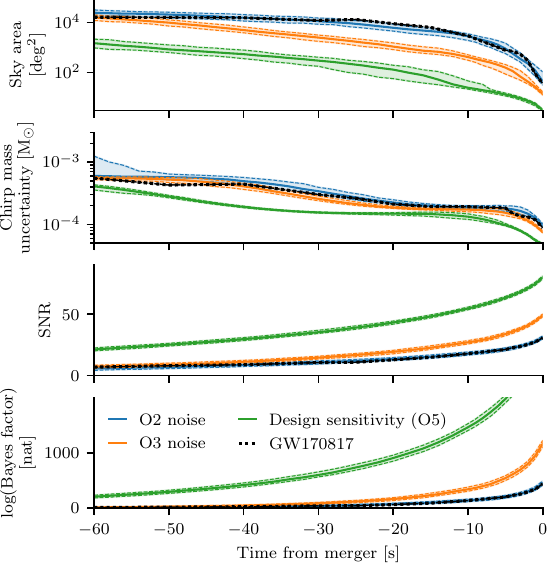}}}
    \end{minipage}
    \hfill
    \begin{minipage}[t]{.49\textwidth}
      \subfloat[\label{fig:bw-comparison}]{\stackinset{l}{0cm}{t}{0cm}{\textbf{b}}{\raggedright\includegraphics[width=\textwidth]{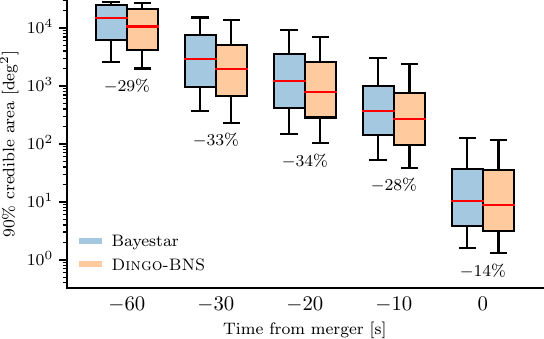}}}\\
      \subfloat[\label{fig:ce-skymaps}]{\stackinset{l}{0cm}{t}{0cm}{\textbf{c}}{\raggedright
      \includegraphics[width=0.32\textwidth]{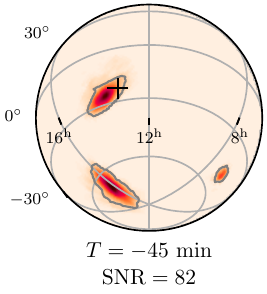}\hspace{0.01\textwidth}
      \includegraphics[width=0.32\textwidth]{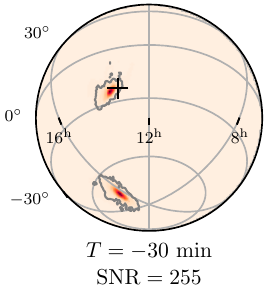}\hspace{0.01\textwidth}
      \includegraphics[width=0.32\textwidth]{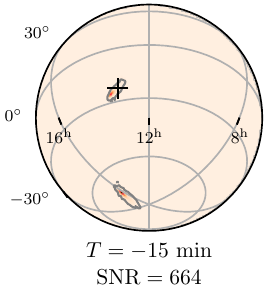}
      }}
    \end{minipage}
    \caption{\textbf{Pre-merger inference with \textsc{Dingo-BNS}.}
    (a) Evolution of pre-merger estimates for GW170817 (black) and GW170817-like simulations injected into different noise levels (colors). We display the 90\% credible sky area, the standard deviation of the chirp mass, the accumulated signal-to-noise and the log Bayes factor comparing the signal and noise models. All of these quantities are inferred with a latency of $\sim1$ second. Dotted lines represent 10th/90th percentiles.
    (b) Sky localization area at 90\% credible level for various premerger times, comparing against Bayestar. The boxplots display the median (red line), quartiles (colored box) and 10th/90th percentiles (whiskers). \textsc{Dingo-BNS} localization is consistently more precise. (c) Premerger sky localization for a GW170817-like event injected into Cosmic Explorer noise, using a minimum frequency of 6~Hz. The black marker indicates the injection coordinates, and gray outline the 90\% credible area.
   }
    \label{fig:results}
\end{figure*}

In this study, we overcome these challenges by leveraging perturbative BNS physics information to simplify and compress the data. However, this simplification requires approximate knowledge of the source itself and is hence valid only over a small portion of the parameter space. We solve this problem using a new algorithm called ``prior-conditioning,'' which enables us to construct networks that can be adapted at inference time to subsets of the prior volume. Our new framework, called \textsc{Dingo-BNS}, makes no (practically relevant) approximation, and takes just \emph{one second} for accurate inference of all 17 BNS parameters (Fig~\ref{fig:dingo-bns}). \textsc{Dingo-BNS} can also infer all of these parameters \emph{minutes before the merger} based on partial inspiral-only information, estimates which can be continuously updated as more data become available (Fig.~\ref{fig:GW170817-injection}).  Near-real-time or pre-merger alerts can then be provided to astronomers, facilitating potential discoveries of precursor and prompt electromagnetic counterparts~\cite{Chaudhary:2023vec,Sridhar:2020uez,Most:2022ojl}.

Our results are faster and more complete than any existing low latency algorithm, with the accuracy of offline parameter estimation codes. Compared to Bayestar, we achieve median reductions in the size of the 90\% credible sky region of 30\% (Fig.~\ref{fig:bw-comparison}). Finally, \textsc{Dingo-BNS} exhibits excellent scaling to longer signals (see \supp), and we demonstrate XG pre-merger inference for signals up to an hour in length (Fig.~\ref{fig:ce-skymaps}).

\sec{\textsc{Dingo-BNS}}For given GW data $d$, we characterize the
source in terms of the posterior probability distribution $p(\theta|d)$ over BNS parameters $\theta$. Parameters include component masses (2), spins (6), orientation, sky position (2), luminosity distance, polarization, time and phase of coalescence, and (in contrast to black holes) tidal deformabilities (2). Following~\cite{Dax:2021tsq}, we use simulated GW datasets to train a density estimation neural network $q(\theta|d)$ (a normalizing flow) to approximate $p(\theta|d)$. Once trained, inference for new $d$ simply requires sampling
$\theta\sim q(\theta|d)$. We obtain asymptotically exact results by augmenting  samples with importance weights using the GW likelihood
function~\cite{LIGOScientific:2019hgc}. This framework, called
\textsc{Dingo-IS}~\cite{Dax:2022pxd}, has been successfully applied to
black hole mergers, however the length of BNS signals renders the na\"ive
transfer of machine learning methods impossible.

\textsc{Dingo-BNS} makes several innovations to tackle this challenge (Fig.~\ref{fig:flowchart}), including using knowledge of specific BNS signal morphology to compress data in a non-lossy way; conditioning the network on the compressor using prior-conditioning; frequency masking based on the pre-merger time and chirp mass; and conditioning on parameter subsets for incorporating multi-messenger information or expectations from nuclear models. The philosophy underlying our approach is that the full BNS problem is too hard for existing neural architectures, so we divide the parameter and data spaces into manageable portions based on known physical information. We then combine all of these variable design choices into a single network that we can instantly tune to the context at hand.

\sec{Data compression and prior conditioning}We
adapt two GW analysis techniques to the SBI context,
heterodyning~\citeheterodyning
to simplify the
data, and multibanding~\citemultibanding to reduce the data
dimension without loss of information. During the long inspiral
period, a BNS signal exhibits a ``chirp,'' with phase evolution (to
leading order in the post-Newtonian expansion~\cite{Blanchet:2013haa}),
\begin{equation}\label{eq:phase-chirp}
    \varphi(f; \Mc) = \frac{3}{128}\left(\frac{\pi G \Mc f}{c^3}\right)^{-5/3},
\end{equation}
where $\Mc = (m_1m_2)^{3/5}/(m_1+m_2)^{1/5}$ is the chirp mass of the
system, with $m_1$, $m_2$ the component masses. Given an approximation
$\Mcp$ to the chirp mass, we heterodyne the (frequency-domain)
data by multiplying by $e^{i\varphi(f; \widetilde\Mc)}$, reducing the number
of oscillations in the signal by several orders of
magnitude~(Fig.~\ref{fig:heterodyned-multibanding}). Given heterodyned
data, we apply multibanding by partitioning the domain into
(empirically-determined) frequency bands, and coarsening the
resolution in higher bands such that the (heterodyned) signal is preserved.

\begin{figure}
  \vspace{-1.5ex}
  \subfloat[\label{fig:prior-conditioning-technical}]{\stackinset{l}{0cm}{t}{0cm}{\textbf{a}}{\hspace{0.03\columnwidth}\includegraphics[width=.98\columnwidth]{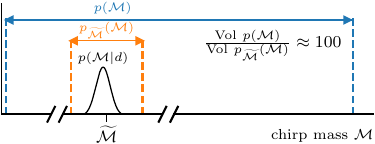}}}\\
  \subfloat[\raggedright b \textnormal{\emph{Training step}}]{
  \begin{minipage}{\columnwidth}
  \smallskip
    \begin{algorithmic}[1]
      \State $\Mcp \sim \hat{p}(\Mcp)$ \Comment{Sample $\Mcp$ from hyperprior}
      \State  $\Mc \sim p_\Mcp(\Mc)$ \Comment{Sample $\Mc$ from restricted prior}
      \State $d \gets d(\Mc)$ \Comment{Simulate data}
      \State $d_\Mcp \xleftarrow{\Mcp} d$ \Comment{Compress data, see \eqref{fig:heterodyned-multibanding}}
      \State Optimize $-\log q(\Mc - \Mcp | d_\Mcp,\Mcp)$
    \end{algorithmic}\smallskip
    \end{minipage}}\\
    \subfloat[\raggedright c \textnormal{\emph{Inference}}]{\begin{minipage}{\columnwidth}
    \smallskip
    \begin{algorithmic}[1]
      \Require  $\Mcp \approx \Mc_\text{true}$ \Comment{Choose approximate $\Mcp$}
      \State $d_\Mcp \xleftarrow{\Mcp} d$ \Comment{Compress data}
      \State $\delta \Mc \sim q(\delta \Mc | d_\Mcp, \Mcp)$ \Comment{Sample $\delta \Mc$ from network}
      \State $\Mc \gets \Mcp + \delta \Mc$
    \end{algorithmic}
    \end{minipage}
    }
  \caption{\label{fig:prior-conditioning}\textbf{Prior conditioning enables event-specific compression.} We train an SBI model simultaneously across a range of priors, each parametrized by a reference chirp mass $\Mcp$. For each (narrow) prior $p_\Mcp(\Mc)$, we apply heterodyning and multibanding compression. This compression simplifies the data distribution that the model must learn and reduces its dimensionality. For simplicity in this presentation, we omit parameters other than the chirp mass.
  }
\end{figure}

Since the compression described requires $\Mcp$ to approximate the chirp mass, it cannot be done across the entire BNS prior volume using a single $\Mcp$ value. \textsc{Dingo-BNS} therefore uses prior-conditioning to restrict to an event-specific prior over which data is compressed. The restricted volume additionally simplifies the density estimation task.
By conditioning on the choice of restriction, prior-conditioning trains a network that is \emph{tunable} to this choice but otherwise applicable over the whole volume  (Fig.~\ref{fig:prior-conditioning}).  
Inference requires an estimate $\Mcp$ of the chirp mass $\Mc$, which can be determined quickly by sweeping across the prior (see \supp).

\algrenewcommand\algorithmicrequire{Require:}

\sec{Frequency masking}In contrast to  past work,
\textsc{Dingo-BNS} also allows for strain frequency series with
varying $f_\text{min}$ and $f_\text{max}$. For a given analysis,
$f_\text{min}$ is chosen based on $\Mcp$ and the segment duration, as the minimum frequency present in the signal in a given GW detector network. This masking is necessary for consistency with frequency-domain waveform models, which assume infinite duration.
Choosing $f_\text{max}$, by contrast, determines the end time of the data stream analyzed to enable pre-merger inference (see \supp).

\sec{Conditioning on parameter subsets}The \textsc{Dingo-BNS} framework (and SBI in general) allows for considerable flexibility in terms of quickly marginalizing over and conditioning on parameters. Conditioning on a parameter allows us to set it to a fixed value, e.g., to incorporate knowledge of that parameter from other sources. 
In our study, we trained \textsc{Dingo-BNS} networks conditioned on the sky position, i.e., we learned $p(\theta \setminus \{\alpha, \delta\} |d, \alpha, \delta)$, where $\alpha$, $\delta$ denote the right ascension and declination, respectively. Such a network allows us to incorporate precise multi-messenger localization to obtain tighter constraints on the remaining parameters, potentially enabling real-time feedback on whether optical candidates should be prioritized for detailed spectroscopy~\cite{LIGOScientific:2019gag}. 
In this way, \textsc{Dingo-BNS} can enable new modes of interaction between GW and electromagnetic observers, potentially transforming how we prioritize and respond to multi-messenger events. We have also explored parameter-conditioning to accelerate offline nuclear equation-of-state analyses (see \supp). 

\sec{Experiments}We generate training data using simulated BNS waveforms (including spin-precession and tidal contributions, but without higher angular multipoles~\cite{Dietrich:2018uni}) with additive stationary Gaussian detector noise. When relevant, networks are also trained with power spectral density (PSD)-conditioning to enable instant tuning to noise levels at the time of an event. At inference time, we validate and correct results using importance sampling, thus guaranteeing their accuracy provided a sufficient effective sample size is obtained~\cite{Dax:2022pxd}. We accelerate the importance sampling step using JAX waveform and likelihood implementations~\cite{Edwards:2023sak,Wong:2023lgb,Wouters:2024oxj}. 

We performed four studies using \textsc{Dingo-BNS}: (a) pre-merger analysis of the first BNS detected, GW170817, as well as equivalent injections (simulated data sets) at varying noise levels; (b) pre-merger analysis of a range of injections in LVK design sensitivity noise; (c) after-merger analysis of the two detected BNS events, GW170817 and GW190425, reproducing published LVK results; and (d) pre-merger analysis of injections in Cosmic Explorer noise (with a minimum frequency of 6 Hz, corresponding to an hour long signal). 
We use the importance sampling efficiency as a primary performance metric, finding average values of 45.8\%, 48.5\%, 31.0\%, and 35.6\% in experiments (a), (b), (c), (d), respectively. 
With these high efficiencies, inference for $10^4$ effective samples takes roughly one second on an H100 GPU (see \supp). Efficiencies are generally higher for pre-merger, likely because the waveform morphology is simplest in the early inspiral. 

\sec{Discussion}Prior conditioning works well for BNS inference, and it could be extended to address further challenges in GW astronomy (e.g., isolation of events from overlapping backgound signals in XG) and other scientific domains. In the future we would like to explore our prior-conditioning approach to data compression for black hole-neutron star systems and low-mass BBHs. This is nontrivial because such systems can emit GWs in higher angular radiation multipoles (i.e., beyond the  $(l, m) = (2,2)$ mode that we assume here), which evolve according to integer multiples of \eqref{eq:phase-chirp}, and so would require an improved heterodyning algorithm to factor out the chirp. Higher modes are not present in BNS signals since the stars are very nearly equal mass.

Another exciting prospect for SBI is a more realistic treatment of detector noise. Indeed, since BNS inspirals have long duration, noise non-stationarities and non-Gaussianities are more likely to manifest. \textsc{Dingo-BNS} currently assumes stationary Gaussian noise and is supplied with an off-source estimate of the PSD. However, by training on realistic detector noise, our approach can in principle learn to fully characterize the noise jointly with the signal, including any deviations from stationarity and Gaussianity. This approach is akin to on-source PSD and glitch modeling~\cite{Cornish:2014kda}, but allows for more general noise and automatically marginalizes over uncertainties. Initial steps in this direction have already been taken for intermediate-mass binary black holes~\cite{Raymond:2024xzj}. Improved noise treatments such as those afforded by SBI will become crucial for reducing systematic error as detectors become more sensitive~\cite{LIGO:2021ppb}. 

Finally, although \textsc{Dingo-BNS} is intended to be used for parameter estimation following a trigger by dedicated search pipelines, its speed opens the possibility to run continuously on all data as they are taken. Either the signal-to-noise ratio or Bayesian evidence time series generated by \textsc{Dingo-BNS} could then be used as a detection statistic, forming an end-to-end detection and parameter estimation pipeline. To implement this would require calibrating these statistics to determine false alarm rates, as well as careful comparisons against existing algorithms to establish efficacy.

\begin{acknowledgments}
\sec{Acknowledgments}We thank S. Buchholz, T. Dent, A. Kofler and S. Morisaki for useful discussions.
This research has made use of data or software obtained from the Gravitational Wave Open Science Center (gwosc.org), a service of the LIGO Scientific Collaboration, the Virgo Collaboration, and KAGRA. This material is based upon work supported by NSF's LIGO Laboratory which is a major facility fully funded by the National Science Foundation, as well as the Science and Technology Facilities Council (STFC) of the United Kingdom, the Max-Planck-Society (MPS), and the State of Niedersachsen/Germany for support of the construction of Advanced LIGO and construction and operation of the GEO600 detector. Additional support for Advanced LIGO was provided by the Australian Research Council. Virgo is funded through the European Gravitational Observatory (EGO), by the French Centre National de Recherche Scientifique (CNRS), the Italian Istituto Nazionale di Fisica Nucleare (INFN) and the Dutch Nikhef, with contributions by institutions from Belgium, Germany, Greece, Hungary, Ireland, Japan, Monaco, Poland, Portugal, Spain. KAGRA is supported by Ministry of Education, Culture, Sports, Science and Technology (MEXT), Japan Society for the Promotion of Science (JSPS) in Japan; National Research Foundation (NRF) and Ministry of Science and ICT (MSIT) in Korea; Academia Sinica (AS) and National Science and Technology Council (NSTC) in Taiwan. M.D.\ thanks the Hector Fellow Academy for support. This work was supported by the German Research Foundation (DFG) through Germany’s Excellence Strategy – EXC- Number 2064/1 – Project number 390727645). The computational work for this manuscript was carried out on the Atlas cluster at the Max Planck Institute for Intelligent Systems in T\"ubingen, Germany, and the Lakshmi and Hypatia clusters at the Max Planck Institute for Gravitational Physics in Potsdam, Germany. V.R. is supported by the UK's Science and Technology Facilities Council grant ST/V005618/1.
\end{acknowledgments}

\bibliography{mybib.bib}

\clearpage
\begin{center}
  \Large
  \textbf{\supp}
\end{center}
\section{Machine learning framework}
The Bayesian posterior $p(\theta|d)=p(d|\theta)p(\theta)/p(d)$ is defined in terms of a prior $p(\theta)$ and a likelihood $p(d|\theta)$. For GW inference, the likelihood is constructed by combining models for waveforms and detector noise. The Bayesian evidence $p(d)$ corresponds to the normalization of the posterior, and it can be used for model comparison.

Our framework is based on neural posterior estimation (NPE)~\cite{papamakarios2016fast,lueckmann2017flexible,greenberg2019automatic}, which trains a density estimation neural network $q(\theta|d)$ to estimate $p(\theta|d)$. We parameterize $q(\theta|d)$ with a conditional normalizing flow~\cite{rezende2015variational,durkan2019neural}.
Training minimizes the loss $L=-\log q(\theta|d)$ across a dataset $(\theta_i,d_i)$ of parameters $\theta_i\sim p(\theta)$ paired with corresponding likelihood simulations $d_i\sim p(d|\theta_i)$. After training, $q(\theta|d)$ serves as a surrogate for $p(\theta|d)$, and inference for any observed data $d_\text{o}$ can be performed by sampling $\theta\sim q(\theta|d_\text{o})$. 
\textsc{Dingo}~\cite{Dax:2021tsq,Dax:2022pxd} uses a group-equivariant formulation of NPE (GNPE~\cite{Dax:2021myb,Dax:2021tsq}), which simplifies GW data by aligning coalescence times in the different detectors. However, this comes at the cost of longer inference times, so we do not use GNPE for \textsc{Dingo-BNS}.

At inference, we correct for potential inaccuracies of $q(\theta|d)$ with importance sampling~\cite{Dax:2022pxd}, by assigning weight $w_i=p(d|\theta_i)p(\theta_i)/q(\theta_i|d)$ to each sample $\theta_i\sim q(\theta_i|d)$. A set of $n$ weighted samples $(w_i,\theta_i)$ corresponds to $n_\text{eff} = \left(\sum_i w_i\right)^2 / \left(\sum_i w_i^2\right)$ \emph{effective} samples from the posterior $p(\theta|d)$. This reweighting enables asymptotically exact results, and the sample efficiency $\epsilon=n_\text{eff}/n$ serves as a performance metric. The normalization of the weights further provides an unbiased estimate of the Bayesian evidence $p(d)=\left(\sum_i w_i\right) / n$. 

Below, we describe in more detail the technical innovations of \textsc{Dingo-BNS} that enable scaling of this framework to BNS signals.

\subsection{Prior conditioning}
\begin{figure*}
    \centering
    \subfloat[\raggedright a\label{fig:supp-mc-scan}]{\includegraphics[width=0.325\textwidth]{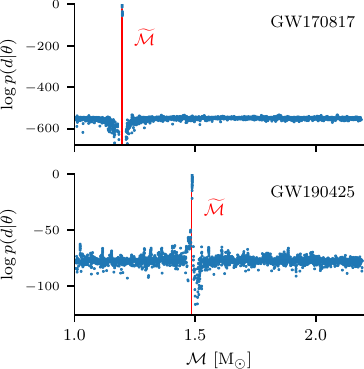}}\hfill
    \subfloat[\raggedright b\label{fig:supp-mc-prior}]{\includegraphics[width=0.33\textwidth]{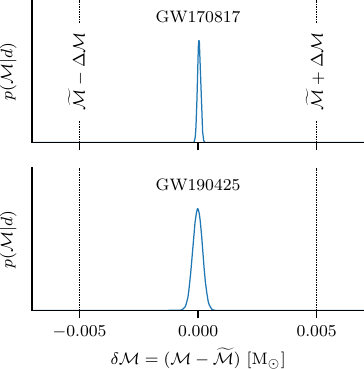}}
    \subfloat[\raggedright c\label{fig:supp-t-scan}]{\includegraphics[width=0.325\textwidth]{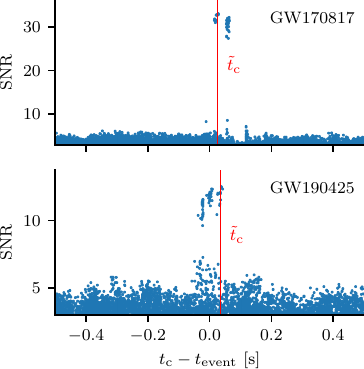}}
    \caption{(a) Log likelihoods generated from a scan over different values of $\Mcp$ with a \textsc{Dingo-BNS} network. The final $\Mcp$ is chosen as the maximum likelihood $\Mc$ (red line; $\Mcp=1.1975~\text{M}_\odot$ for GW170817, $\Mcp=1.4868~\text{M}_\odot$ for GW190425). (b) Posterior marginal $p(\Mc|d)$. The prior (dashed lines) determined by the scan from (a) fully covers the marginal. (c) A combined scan over $\Mc$ and $t_\text{c}$ successfully identifies GW170817 (with $\hat t_\text{c}=1187008882.43$) and GW190425 (with $\hat t_\text{c}=1240215503.04$).}
    \label{fig:scan}
\end{figure*}

An NPE model $q(\theta|d)$ estimates the posterior $p(\theta|d)$ for a fixed prior $p(\theta)$. Choosing a broad prior enhances the general applicability of the NPE model, but it also implies worse tuning to specific events (for which smaller priors may be sufficient). 
This is a general trade-off in NPE, but it is particularly dramatic for BNS inference, where typical events constrain the chirp mass to $\sim 10^{-3}$ of the prior volume. 
Thus, for an individual BNS event, a tight chirp mass prior would have been sufficient (Fig.~\ref{fig:supp-mc-prior}) and moreover would have enabled effective heterodyning~\cite{Cornish:2010kf,Cornish:2021lje,Zackay:2018qdy}; but in order to cover generic BNS events, we need to train the NPE network with a large prior (see Tab.~\ref{tab:priors}).
We resolve this trade-off with a new technique called prior conditioning. The key idea is to train an NPE model with \emph{multiple different} (restricted) priors simultaneously. On each of these priors, we are allowed to apply an independent transformation to the data, which we use to heterodyne the GW strain with respect to the approximate chirp mass. Training a prior-conditioned model requires hierarchical sampling
\begin{align}
    \theta\sim p_\rho(\theta),~\rho\sim \hat p(\rho),
\end{align}
where $p_\rho(\theta)$ is a prior family parameterized by $\rho$ and $\hat p(\rho)$ is a corresponding hyperprior. We additionally condition the NPE model $q(\theta|d,\rho)$ on $\rho$. This model can then perform inference for any desired prior $p_\rho(\theta)$, by simply providing the corresponding $\rho$. This effectively amortizes the training cost over different choices of the prior.

We apply prior conditioning for the chirp mass $\Mc$, using a set of priors $p_\Mcp(\Mc) = U_{m_1,m_2}(\Mcp - \Delta\Mc, \Mcp + \Delta\Mc)$. Here, $U_{m_1,m_2}(\Mc_\text{min}, \Mc_\text{max})$ denotes a distribution over $\Mc$ with support $[\Mc_\text{min}, \Mc_\text{max}]$, within which component masses $m_1$, $m_2$ are uniformly distributed. We use fixed $\Delta \Mc = 0.005~\text{M}_\odot$ and choose a hyperprior $\hat p(\Mcp)$ covering the expected range of $\Mc$ for LVK detections of BNS (see Tab.~\ref{tab:priors}). As $\Delta\Mc$ is small, $\Mcp$ is a good approximation for any $\Mc$ within the restricted prior $p_\Mcp(\Mc)$ and we can thus use $\Mcp$ for heterodyning. The resulting model $q(\theta|d_{\Mcp}, \Mcp)$ can then perform inference with event-optimized heterodyning and prior (via choice of appropriate $\Mcp$), but is nevertheless applicable to the entire range of the hyperprior. 

Inference results are independent of $\Mcp$ as long as the posterior $p(\Mc|d)$ is fully covered by $[\Mcp - \Delta\Mc, \Mcp + \Delta\Mc]$. For BNS, $p(\Mc|d)$ is typically tightly constrained and we can use a coarse estimate of $\Mc$ for $\Mcp$. This can either be taken from a GW search pipeline or rapidly computed from $q(\theta|d_{\Mcp}, \Mcp)$ itself by sweeping the hyperprior (see below). Note that for shorter GW signals from black hole mergers, $p(\Mc|d)$ is generally less well constrained. Transfer of prior conditioning would thus require larger (and potentially flexible) values of $\Delta\Mc$. Alternatively, the prior range can be extended at inference time by iterative Gibbs sampling of $\Mc$ and $\Mcp$, similar to the GNPE algorithm~\cite{Dax:2021tsq,Dax:2021myb}. 

Prior conditioning is a general SBI technique that enables choice of prior at inference time. This can also be achieved with sequential NPE~\cite{papamakarios2016fast,lueckmann2017flexible,greenberg2019automatic,deistler2022truncated}. However, in contrast to prior condditioning, these techniqes require simulations and retraining for each observation, resulting in more expensive and slower inference. We here use prior conditioning with priors of fixed width for the chirp mass, and optional additional conditioning on fixed values for other parameters (corresponding to Dirac delta priors). Extension to more complicated priors and hyperpriors is straightforward.

\subsection{Independent estimation of chirp mass and merger times}
Running \textsc{Dingo-BNS} requires an initial estimate of the chirp mass $\Mc$ (to determine $\Mcp$ for the network) and the merger time $t_\text{c}$ (to trigger the analysis). 
Matched filter searches can identify the presence of a compact binary signal signal and its chirp mass and merger time in low-latency~\cite{Finn:1992wt,Nitz:2018rgo, Cannon:2020qnf,Adams:2015ulm,Chu:2020pjv}. Specialized “early warning” searches designed to produce output before the coalescence can further provide a rough indication of sky position and distance~\cite{Sachdev:2020lfd, Nitz:2020vym, Kovalam:2021bgg}. 
When available, output of such pipelines can be used to trigger a \textsc{Dingo} analysis and provide estimates for $\Mc$ and $t_\text{c}$.

We here describe an alternative independent approach of obtaining these parameters, using only the trained \textsc{Dingo-BNS} model. We compute $\Mcp$ by sweeping the entire hyperprior $\hat p(\Mcp)=U_{m_1,m_2}(\Mcp_\text{min}, \Mcp_\text{max})$. Specifically, we run \textsc{Dingo-BNS} with a set of prior centers 
\begin{equation}
    \Mcp_i = \Mcp_\text{min} + i\cdot\Delta \Mc,~i\in[0,(\Mcp_\text{max}-\Mcp_\text{min})/\Delta \Mc].
\end{equation}
The inference models in this study are trained with hyperprior ranges of up to $[1.0, 2.2]\,\text{M}_\odot$. For $\Delta \Mc = 0.005~\text{M}_\odot$, we can thus cover the entire global chirp mass range using 241 (overlapping) local priors. We run \textsc{Dingo-BNS} for all local priors $\Mcp_i$ in parallel, with 10 samples per $\Mcp_i$. This requires \textsc{Dingo-BNS} inference of only a few thousand samples, which takes less than one second. We use the chirp mass $\Mc$ of the maximum likelihood sample as the prior center $\Mcp$ for the analysis (Fig.~\ref{fig:supp-mc-scan}). Note that the exact choice of $\Mcp$ does not matter, as long as the inferred posterior is fully covered by $[\Mcp - \Delta\Mc, \Mcp + \Delta\Mc]$ (Fig.~\ref{fig:supp-mc-prior}).

The merger time $t_\text{c}$ can be inferred by continuously running this $\Mcp$ scan on the input data stream, sliding the $t_\text{c}$ prior in real time over the incoming data. 
With inference times of one second, continuous analysis could be achieved on just a few parallel computational nodes, constantly running on the input data stream. 
Event candidates can then be identified by analyzing the SNR, triggering upon exceeding some defined threshold (Fig.~\ref{fig:supp-t-scan}). This scan could be performed at an arbitrary (but fixed) time prior to the merger.

This scan successfully estimates $\Mc$ and $t_\text{c}$ for both real BNS events (Fig~\ref{fig:scan}). However, we have not tested this at a large scale on detector noise to compute false alarm rates, as \textsc{Dingo-BNS} is primarily intended for parameter estimation. Existing search and early warning pipelines are likely more robust for event identification, in particular in the presence of non-stationary detector noise. 

\begin{figure*}
    \centering
    \subfloat[\raggedright a\label{fig:multibanding-f}]{\includegraphics[width=0.36\textwidth]{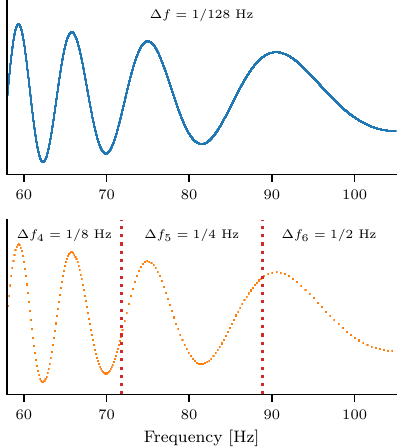}}
    \hfill
    \subfloat[\raggedright b\label{fig:multibanding-bins}]{\includegraphics[width=0.36\textwidth]{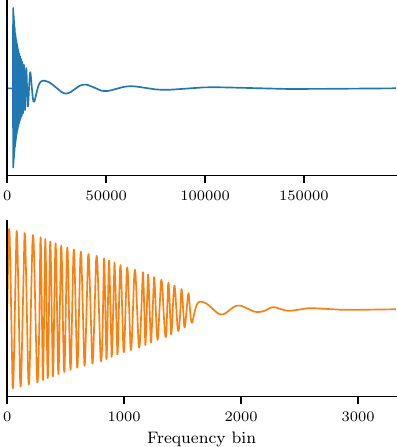}}
    \hfill
    \subfloat[\raggedright c\label{fig:multibanding-nodes}]{
    \begin{tabular}{r@{\hskip 0.05in}|@{\hskip 0.05in}r@{\hskip 0.1in}r}
        \hline\hline
        & LVK & CE \\
        $\Delta f_i~[\text{Hz}]$ & $\hat{f}_i~[\text{Hz}]$ & $\hat{f}_i~[\text{Hz}]$ \\ 
        \hline
        $1 / 8192$ &  & 5 \\
        $1 / 4096$ &  & 5 \\
        $1 / 2048$ &  & 5 \\
        $1 / 1024$ &  & 7 \\
        $1 / 512$ &  & 10 \\
        $1 / 256$ &  & 13 \\
        $1 / 128$ & 19 & 18 \\
        $1 / 64$ & 25 & 24 \\
        $1 / 32$ & 34 & 32 \\
        $1 / 16$ & 44 & 42 \\
        $1 / 8$ & 57 & 54 \\
        $1 / 4$ & 72 & 68 \\
        $1 / 2$ & 89 & 84 \\
        $1$ & 420 & 364 \\
        $2$ &  & 765 \\
        \hline\hline
    \end{tabular}
    }
    \caption{Frequency multibanding. (a) The period of (heterodyned) GW signals decreases with increasing frequency. The native frequency resolution (blue) thus oversamples the signal at high frequencies. Frequency multibanding (band boundaries indicated by dotted red lines) adapts to the signal variation, decreasing the resolution at higher frequencies (orange). 
    (b) The multibanded domain therefore requires fewer frequency bins, and the signal variation is more homogeneous across bins. 
    (c) Multibanded frequency domain partitions for LVK ($f_\text{min} = 19.4~\text{Hz}$, compression factor $\sim 60$) and CE ($f_\text{min} = 5\,\text{Hz}$, compression factor $\sim 650$) experiments.
    We use a smaller chirp mass prior for the CE experiments (Tab.~\ref{tab:priors}), which allows a slightly coarser resolution compared to LVK (corresponding to lower $\hat{f}_i$). The first two bands for CE are skipped entirely, which is a consequence of the reduced signal variation with heterodyning.
    }
    \label{fig:frequency-multibanding}
\end{figure*}

\subsection{Frequency multibanding}
Although the native resolution of a frequency series is determined by the duration $T$ of the corresponding time series ($\Delta f=1/T$), we can average adjacent frequency bins wherever the signal is roughly constant. This enables data compression with only negligible loss of information. We here employ frequency multibanding, which divides the frequency range $[f_\text{min}, f_\text{max}]$ into $N$ bands of decreasing resolution. Frequency band $i$ covers the range $[\hat{f}_i, \hat{f}_{i+1})$ with $\Delta f_i = 2^i \Delta f_0$, where $\hat{f}_0=f_\text{min}$, $\hat{f}_N=f_\text{max}$ and $\Delta f_0$ is the native resolution of the frequency series. Within a band $i$, the multibanded domain thus compresses the data by a factor of $2^i$ (Fig.~\ref{fig:frequency-multibanding}), which is achieved by averaging $2^i$ sequential bins from the original frequency series (``decimation''). To achieve optimal compression, we empirically choose the smallest possible nodes $\hat{f}_i$ for which GW signals are still fully resolved. Specifically, we simulate a set of $10^3$ heterodyned GW signals and demand that every period of these signals is covered by at least 32 bins in the resulting multibanded frequency domain. This is done before generating the training dataset, and the multibanded domain then remains fixed during dataset generation and training. The optimized resolution achieves compression factors between 60 and 650 (Fig.~\ref{fig:multibanding-nodes}). Care needs to be taken that our approximations are valid in the presence of detector noise. We now investigate how multibanding affects \emph{data simulation} (for training) and the \emph{likelihood} (for importance sampling).

\sec{Data simulation}GW data is simulated as the sum of a signal and detector noise, $d=h(\theta)+n$. The detector noise in frequency bin $j$ is given by
\begin{equation}
    \label{eq:base_std}
    n_j \sim \mathcal{N}(0,\sigma\sqrt{S}_j),
    \quad \sigma=\sqrt{\frac{w}{4 \Delta f}},
\end{equation}
where $S$ denotes the detector noise PSD and $\sigma$ takes into account the frequency resolution and the Tukey window factor $w$. Note that $n$ is a complex frequency series, which we ignore in our notation, as the considerations here hold for real and imaginary part individually. It is conventional to work with whitened data
\begin{equation}
    d_j^\text{w} = h_j^\text{w}(\theta) + n_j^\text{w} = \frac{h_j(\theta) + n_j}{\sqrt{S_j}},
\end{equation}
in which case $n_j^\text{w} \sim \mathcal{N}(0,\sigma)$.

We convert to multibanded frequency domain by averaging sets of $N_i = 2^i$ bins,  
\begin{equation}
\begin{aligned}
\overline{d_{j}^\text{w}} &= \frac{1}{N_i} \sum_{k=m_j}^{m_j+N_i-1} \left(h_k^\text{w} + n_k^\text{w} \right) = \overline{h_{j}^\text{w}} + \overline{n_{j}^\text{w}},
\end{aligned}
\end{equation}
where $j$ denotes the bin in the multibanded domain, $m_j$ denotes the starting index of the decimation window for $j$ in the native domain, and $i$ indexes the frequency band associated with $j$. Since $\overline{n_{j}^\text{w}}$ is an average of $N_i$ Gaussian random
variables with standard deviation $\sigma$, it follows that $\overline{n_{j}^\text{w}}$ is also Gaussian with standard deviation
\begin{equation}
    \sigma_i = \sigma/\sqrt{N_i}=\sqrt{\frac{w}{4 \Delta f N_i}}=\sqrt{\frac{w}{4 \Delta f_i}}. 
\end{equation}
We can thus simulate the detector noise directly in the multibanded domain, by updating $\sigma\rightarrow\sigma_i$, corresponding to $\Delta f\rightarrow\Delta f_i$. 
For the whitened signal we find
\begin{equation}
    \overline{h^\text{w}_{j}} 
    = \frac{1}{N_i} \sum_{k=m_j}^{m_j+N_i-1} \frac{h_{k}}{\sqrt{S_k}} 
    \approx \overline{h_{j}} \sum_{k=m_j}^{m_j+N_i-1}  \frac{1}{\sqrt{S_k}},
\end{equation}
assuming an approximately constant signal $h$ within the decimation window, $\overline{h_{j}} \approx h_k,\forall k\in [m_j, m_j + N_i - 1]$. For frequency-domain waveform models, We can thus directly compute the signal $\overline{h}_j$ in the multibanded domain by simply evaluating the model at frequencies $\overline{f_j}$. For whitening, we replace $1/\sqrt{S}\rightarrow \overline{1/\sqrt{S}}$.

In summary, we can directly generate BNS data in the multibanded frequency domain, by (1) updating the noise standard deviation according to the multibanded resolution, (2) appropriately decimating noise PSDs and (3) computating signals and noise realizations in the compressed domain. 
These operations are carefully designed to be consistent with data processing of real BNS observations, which for \textsc{Dingo-BNS} are first whitened in the native domain and then decimated to the multibanded domain.
This process relies on the assumption that signals are constant within decimation windows, and we ensure that this is (approximately) fulfilled when determining the multibanded resolution. 
Indeed, for signals generated directly in the multibanded domain we find mismatches of at most $\sim 10^{-7}$ when comparing to signals that are properly decimated from the native domain.

\sec{Likelihood evaluations}We also use frequency multibanding to evaluate the likelihood for importance sampling. The standard Whittle likelihood used in GW astronomy~\cite{LIGOScientific:2019hgc} reads
\begin{equation}
    \log p(d| \theta) = -\frac{1}{2} \sum_{k} \frac{|d_k^\text{w} - h_k^\text{w}(\theta)|^2}{\sigma^2},
\end{equation}
up to a normaliztion constant. The sum extends over all bins $k$ in the native frequency domain. Assuming a constant signal (as above) and PSD within each decimation window, we can directly compute the likelihood in the multibanded domain
\begin{equation}
    \log p(d| \theta) \approx -\frac{1}{2} \sum_{j} \frac{|\overline{d_j^\text{w}} - \overline{h_j^\text{w}}(\theta)|^2}{\sigma_{i(j)}^2}.
\end{equation}
The assumptions are not exactly fulfilled in practice; for additional corrections see~\cite{Morisaki:2021ngj}. For importance sampling, we can always evaluate the exact likelihood in the native frequency domain instead. In this case, the result is no longer subject to any approximations, even if the \textsc{Dingo-BNS} proposal is generated with a network using multibanded data. With the full likelihood for GW170817, we find a sample efficiency of $11.0\%$ with an inference time of $13~\text{seconds}$ for 50,000 samples. The deviation from the result obtained with the multibanded likelihood is negligible (Jensen-Shannon divergence less than $5\cdot 10^{-4}~\text{nat}$ for all parameters). This demonstrates that use of the multibanded resolution has no practically relevant impact on the results. 

\subsection{Frequency masking}
\begin{figure}
    \centering
    \includegraphics[width=\columnwidth]{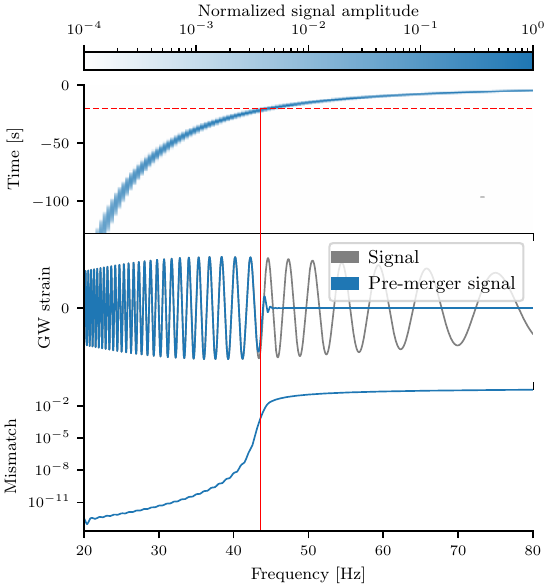}
    \caption{Time-domain truncation of BNS signals at time $t_\text{max}$ (red dashed line) before the merger can be approximated by truncation at a corresponding maximum frequency (red solid line) in frequency domain. Below frequency $f_\text{max}(t, \Mc)$, the truncated signal (blue in center panel) matches the original signal (gray). Above $f_\text{max}(t, \Mc)$, the amplitude of the truncated signal quickly approaches zero. We determine $f_\text{max}(t, \Mc)$ empirically, by allowing mismatches between truncated and original signals of at most $10^{-3}$ (lower panel). Analogously, truncation for $t < t_\text{min}$ can be achieved by imposing a minimum frequency cutoff $f_\text{min}(t, \Mc)$.}
    \label{fig:frequency-masking}
\end{figure}

Since the GW likelihood (and our framework) use frequency domain, but data are taken in time domain, it is necessary to convert data by windowing and Fourier transforming.
However, frequency domain waveform models assume infinite time duration, leading to inconsistencies with finite time segments $[t_\text{min}, t_\text{max}]$. 
As the frequency evolution of the inspiral is tightly constrained by the chirp mass $\Mc$, we can compute boundaries $f_\text{min}(t_\text{min}, \Mc)$ and $f_\text{max}(t_\text{max}, \Mc)$, such that signals are not corrupted by finite-duration effects within $[f_\text{min}, f_\text{max}]$, and are negligibly small outside of that range (Fig.~\ref{fig:frequency-masking}). 

We approximate the lower bound $f_\text{min}(t_\text{min}, \Mc)$ using the leading order in the post-Newtonian relationship between time and frequency,
\begin{equation}
    f_\text{0PN}(t, \Mc) = \frac{1}{8\pi}\left(\frac{-t}{5}\right)^{-3/8} \left(\frac{G\Mc}{c^3}\right)^{-5/8}.
\end{equation}
For a network designed for fixed data duration $T$, we set $f_\text{min}(T, \Mc) = f_\text{0PN}(-T, \Mc) + f_\text{buffer}$ (we use $f_\text{buffer} = 1\,\text{Hz}$ for LVK and $f_\text{buffer} = 0.5\,\text{Hz}$ for XG setups). 

For the upper bound, we found that $f_\text{0PN}(t, \Mc)$ is not sufficiently accurate. Instead, we determine $f_\text{max}(t,\Mc)$ empirically by simulating a set of signals (with parameters $\theta\sim p(\theta)$), and computing mismatches between signals with and without truncation at $t > t_\text{max}$. For a given set of simulations, we choose $f_\text{max}(t,\Mc)$ as the highest frequency for which all mismatches are at most $10^{-3}$. To avoid additional computation at inference time, we cache the results in a lookup table for $f_\text{max}(t,\Mc)$. 

Both bounds depend on the chirp mass $\Mc$, and the upper bound additionally depends on the pre-merger time. To enable inference for arbitrary configurations, we train a single network with variable frequency bounds. During training, we compute $f_\text{min}(T, \Mcp)$ with the center $\Mcp$ of the local chirp mass prior. The upper frequency bound $f_\text{max}$ is sampled randomly (uniform in frequency bins of the multibanded frequency domain) to allow for arbitrary pre-merger times. Data outside of $[f_\text{min}, f_\text{max}]$ is zero-masked. 

\subsection{Equation-of-state likelihood}

\begin{figure}
    \centering
    \includegraphics[width=\columnwidth]{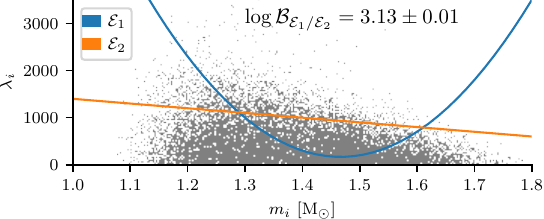}
    \caption{Neutron-star EOS imply a functional relation $\lambda_i=\lambda_i(m_i)$ between tidal parameters $\lambda_i$ and component masses $m_i$. The likelihood $p(d|\mathcal{E})$ for an EOS $\mathcal{E}$ given the GW data $d$ requires integrating the posterior $p(\theta|d)$ along the corresponding hyperplane. No posterior samples (gray) will be exactly on the corresponding hyperplane (exemplary coloured lines), hence the standard Bayesian inference techniques are not directly applicable \cite{Vivanco2019}. \textsc{Dingo-BNS} provides various possibilities to directly compute this quantity, enabling comparison of different EOS in terms of the Bayes factors $\mathcal{B}$. 
    }
    \label{fig:eos}
\end{figure}

A nuclear equation of state (EOS) implies a functional relationship between neutron star masses $m_i$ and tidal deformabilities $\lambda_i$. The likelihood $p(d|\mathcal{E})$ for a given EOS $\mathcal{E}$ and data $d$ can be computed by integrating the GW likelihood along the hyperplane defined by the EOS constraint $\lambda_i = \lambda^\mathcal{E}_i(m_i)$,
\begin{equation}\label{eq:EOS-integration}
    \begin{split}
        p(d|\mathcal{E}) &= \int p(d|\theta)p(\theta)\delta(\lambda_i -\lambda^\mathcal{E}_i(m_i))\, \text{d}\theta\\
        &= \int p(d|m_1,m_2,\lambda^\mathcal{E}_1(m_1),\lambda^\mathcal{E}_2(m_2)) \,\text{d} m_1 \text{d} m_2.
    \end{split}
\end{equation}
Here, $p(d|m_1,m_2,\lambda_1,\lambda_2)$ is the Bayesian evidence of $d$ conditional on $(m_1,m_2,\lambda_1,\lambda_2)$. To calculate \eqref{eq:EOS-integration} using Monte Carlo integration, it is necessary to repeatedly evaluate the integrand, which is extremely expensive using traditional methods (e.g., nested sampling).

With \textsc{Dingo-BNS}, there are two fast ways to evaluate the integrand, using either a conditional or a marginal network: (1) a marginal network $q(m_1,m_2,\lambda_1,\lambda_2|d)$ directly provides an unnormalized estimate of the conditional evidence $p(d|m_1,m_2,\lambda_1,\lambda_2)$ (sufficient for model comparison, but not subject to our usual accuracy guarantees); or (2) a conditional network $p(\theta|d;m_1,m_2,\lambda_1,\lambda_2)$ provides the normalized conditional evidence via importance sampling (including accuracy guarantees). Option (1) allows for $10^5$ evaluations per second, whereas option (2) only allows for $10^3$ assuming $10^2$ weighted samples per evaluation. By combining (1) and (2), we can achieve speed and accuracy, by using the marginal network (1) to define a histogram proposal for Monte Carlo integration with the integrand from (2). We test this on GW170817 data with two polynomial EOS constraints $\lambda = \lambda^\mathcal{E}(m)$ (Fig.~\ref{fig:eos}), finding good sample efficiencies of $\sim 50\%$, small uncertainties $\sigma_{\log p(d|\mathcal{E})}\sim 0.01$ and computation times of $1-3~\text{s}$ for the integral~\eqref{eq:EOS-integration}. Alternatively, the proposal could also be generated with a network $q(m_1,m_2|d)$, which additionally marginalizes over $\lambda_i$. Finally, for a parametric EOS, a \textsc{Dingo-BNS} network could be conditioned on EOS parameters, allowing for direct EOS inference. This variety of approaches emphasizes the flexibility of SBI for EOS inference.

\subsection{Related work}
Machine learning for GW astronomy is an active area of research~\cite{Cuoco:2020ogp}. Several studies explore machine learning inference for black hole mergers~\cite{Gabbard:2019rde,Green:2020hst,Delaunoy:2020zcu,Green:2020hst,Green:2020dnx,Dax:2021tsq,Williams:2021qyt,Dax:2022pxd,Alvey:2023naa,Crisostomi:2023tle,Bhardwaj:2023xph,Wildberger2023flow,Kolmus:2024scm}. There have also been applications specifically to BNS inference, notably the GW-SkyLocator algorithm~\cite{Chatterjee:2022dik}, which estimates the sky position using the SNR time series (similar to Bayestar), and JIM~\cite{Wong:2023lgb,Wouters:2024oxj}, which uses hardware acceleration and machine learning to speed up conventional samplers and achieve full inference in 25 minutes. The ASTREOS framework uses machine learning for BNS equation-of-state inference~\cite{McGinn:2024nkd}. Pre-merger localization with conventional techniques has also been explored in~\cite{Hu:2023hos}.

\section{Experimental details}

\begin{table}
    \centering
    \begin{tabular}{lrr}
        \hline\hline
        & LVK & CE \\\hline
        $\Mc~[\text{M}_\odot]$ & $[1.0, 2.2]$ & $[1.15, 1.25]$ \\
        $m_1$ & [1.0, 3.2] & [0.95, 2.4] \\
        $m_2$ & [1.0, 2.0] & [0.95, 2.4] \\
        $a_{1,2}$ & [0, 0.05] & [0, 0.05] \\
        $\lambda_1$ & [0, 5000] & [0, 5000] \\
        $\lambda_2$ & [0, 10000] & [0, 10000] \\
        $d_\text{L}~[\text{Mpc}]$ & [10, 100] & [20, 50] / [1000, 2000] \\
        $t_\text{c}~[\text{s}]$ & [-0.1, 0.1] / [-0.03, 0.03] & [-1,1] / [-0.03, 0.03] \\
        \hline\hline
    \end{tabular}
    \caption{Training priors for chirp mass $\Mc$, component masses $m_{1,2}$, spin magnitudes $a_{1,2}$, tidal deformabilities $\lambda_{1,2}$, luminosity distance $d_\text{L}$ and merger time $t_\text{c}$. All priors are uniform, except for chirp mass, which is sampled uniform in component masses. 
    At inference, $d_\text{L}$ can be reweighted to the standard prior (uniform in comoving volume). 
    For $t_\text{c}$ we use a broader prior for pre-merger inference than for full inference (separated by ``/'' symbol) to account for higher uncertainties. 
    LVK priors are chosen to cover expected LVK BNS detections. CE priors for $\Mc$ and $d_\text{L}$ are reduced compared to LVK to decrease the computational cost of training. 
    Priors for parameters not displayed here are standard.
    }
    \label{tab:priors}
\end{table}

For our experiments, we train \textsc{Dingo-BNS} networks using the hyperparameters and neural architecture~\cite{he2015deep,durkan2019neural} from Ref.~\cite{Dax:2021tsq}, with a slightly larger embedding network. For the LVK experiments, we use a dataset with $3\cdot 10^7$ training samples and train for 200 epochs, for CE we use $6\cdot 10^7$ training samples and train for 100 epochs. We use three detectors for LVK (LIGO-Hanford, LIGO-Livingston, and Virgo) and two detectors for CE (primary detector at location of LIGO-Hanford, secondary detector at location of LIGO-Livingston). The networks are trained with the priors displayed in Tab.~\ref{tab:priors}.

In the first experiment, we evaluate \textsc{Dingo-BNS} models on 200 simulated GW datasets, generated using a fixed GW signal with GW170817-like parameters and simulated LVK detector noise. We use noise PSDs from the second (O2) and third (O3) LVK observing runs as well as LVK design sensitivity. For each noise level, we train one pre-merger network ($f\in[23, 200]~\text{Hz}$) and one network for inference with the full signal, including the merger ($f\in[23, 1024]$). The latter network is only used for after-merger inference, as we found that separation into two networks improves the performance. The pre-merger network is trained with frequency masking with the masking bound $f_\text{max}$ sampled in range $[28, 200]~\text{Hz}$, enabling inference up to 60~seconds before the merger. 

\begin{figure*}
    \centering
    \includegraphics[width=0.32\textwidth]{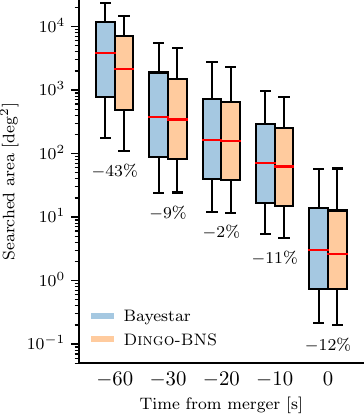}
    \includegraphics[width=0.32\textwidth]{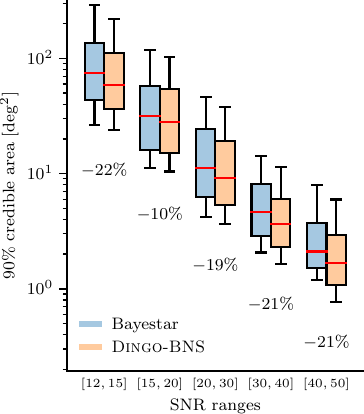}
    \includegraphics[width=0.32\textwidth]{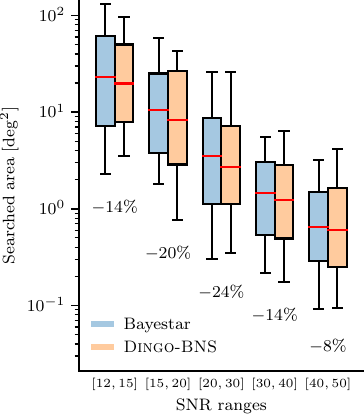}
    \caption{Localization comparison between Bayestar and Dingo-BNS, in terms of the 90\% credible area and the searched area. The comparisons according to SNR are based only on results after the merger.}
    \label{fig:bayestar-supplemental}
\end{figure*}

In the second experiment, we analyze 1000 simulated GW datasets, with GW signal parameters randomly sampled from the prior (Tab.~\ref{tab:priors}; $\Mc$ prior reduced to range $[1.0, 1.5]~\text{M}_\odot$ and $d_\text{L}$ prior reweighted to a uniform distribution in comoving volume) and with design sensitivity noise PSDs. We again train one pre-merger network ($f\in[19.4, 200]~\text{Hz}$) and one after-merger network ($f\in[19.4, 1024]~\text{Hz}$). The pre-merger network is trained with frequency masking with the masking bound $f_\text{max}$ sampled in range $[25, 200]~\text{Hz}$, enabling inference up to 60~seconds before the merger for $\Mc \leq 1.5~\text{M}_\odot$. Both networks are additionally trained with lower frequency masking, with $f_\text{min}(\Mcp)$ determined as explained above, ensuring an optimal frequency range for any chirp mass. For each \textsc{Dingo-BNS} result, we generate a skymap using a kernel density estimator implemented by \verb|ligo.skymap|~\cite{Singer:ligoskymap}. For the sky localization comparison between \textsc{Dingo-BNS} and Bayestar, we run Bayestar based on the GW signal template generated with the maximum likelihood parameters from the \textsc{Dingo-BNS} analysis. We note that Bayestar is designed as a low-latency pipeline and typically run with (coarser) parameter estimates from search templates. Therefore, the reported Bayestar runs may deviate slightly from the realistic LVK setup. However, our results are consistent with Ref.~\cite{Morisaki:2023kuq}, which also found a $\sim 30\%$ precision improvement over Bayestar localization (using LVK search triggers). Both, \textsc{Dingo-BNS} and Ref.~\cite{Morisaki:2023kuq}, perform full Bayesian BNS inference and should therefore have identical localization improvements over Bayestar (assuming ideal accuracy, which for \textsc{Dingo-BNS} is validated with consistently high importance sampling efficiency). Differences to the localization comparison in Ref.~\cite{Morisaki:2023kuq} are thus primarily attributed to different configurations for Bayestar and slightly different injection priors. Additional results for the localization comparison are shown in Fig.~\ref{fig:bayestar-supplemental}.

\begin{figure}
  \includegraphics[width=\columnwidth]{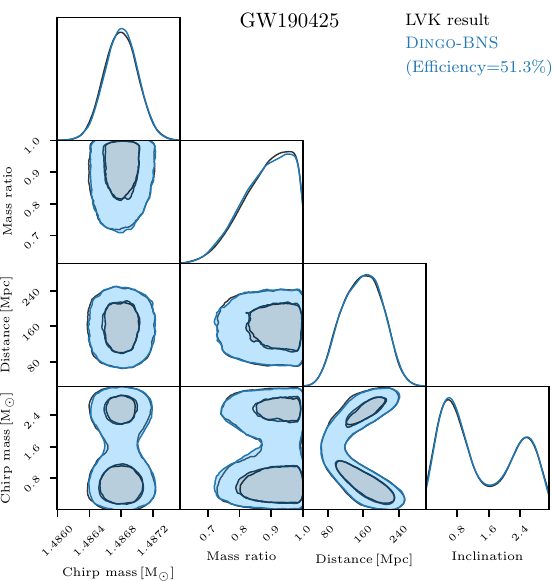}
  \caption{\label{fig:GW190425}Inference results for GW190425, displaying the 50\% and 90\% credible regions for the 2D marginals. \textsc{Dingo-BNS} shows good agreement with the public LVK result.}
\end{figure}

In the third experiment, we reproduce the public LVK results for GW170817~\cite{TheLIGOScientific:2017qsa,Abbott:2018wiz} and GW190425~\cite{Abbott:2020uma} with \textsc{Dingo-BNS}. We use the same priors and data settings as the LVK, but we do not marginalize over calibration uncertainty. We find good sample efficiencies for both events (10.8\% for GW170817 and 51.3\% for GW190425) and good agreement with the LVK results (Fig.~\ref{fig:GW190425}). The LVK results use detector noise PSDs generated with BayesWave~\cite{Cornish:2014kda}, which are not available prior to the merger. For our pre-merger analysis of GW170817 in the main part we thus use a PSD generated with the Welch method. 
The GW170817 signal overlapped with a loud glitch in the LIGO-Livingston detector~\cite{TheLIGOScientific:2017qsa}, and we use the glitch subtracted data provided by the LVK in our analyses. Since such data would not be available prior to the merger, pre-merger inference of BNS events overlapping with glitches would in practice also require fast glitch mitigation methods.

In the fourth experiment, we analyze simulated CE data using the anticipated noise PSDs for the primary and secondary detectors. We train a \textsc{Dingo-BNS} network for pre-merger inference with $f\in [6, 11]~\text{Hz}$, with the upper frequency masking bound $f_\text{max}$ sampled in range $[7, 11]~\text{Hz}$. This supports a signal length of 4096~seconds, with pre-merger inference between 45 and 15 minutes prior to the merger. We inject signals with GW170817-like parameters for distance, masses and inclination, to investigate how well GW170817-like event can be localized in the CE detector. We also train a network on the full frequency range $[6, 1024]~\text{Hz}$ for after-merger inference, with a reduced distance prior to control the SNR (Tab.~\ref{tab:priors}).

\subsection{Sample efficiencies}

\begin{figure*}
    \centering
    \subfloat[\raggedright a\label{fig:efficiency-GW170817-like}]{
    \includegraphics[width=0.30\textwidth]{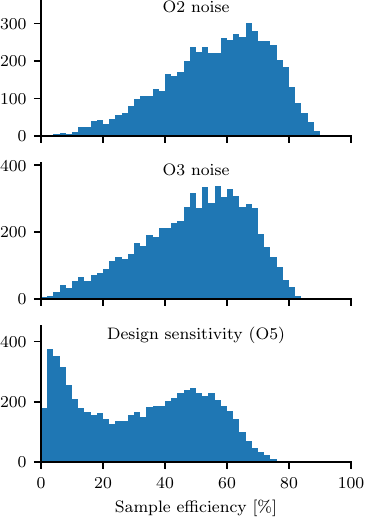}}
    \hfill
    \subfloat[\raggedright b\label{fig:efficiency-DS}]{
    \includegraphics[width=0.30\textwidth]{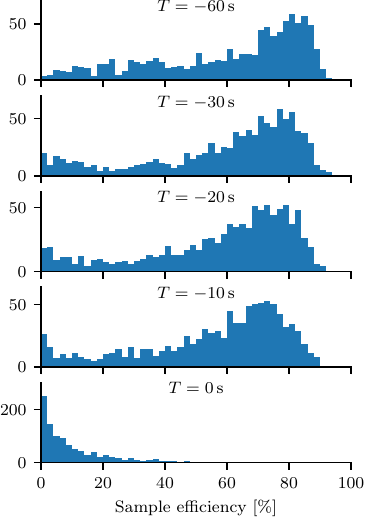}}
    \hfill
    \subfloat[\raggedright c\label{fig:efficiency-CE}]{
    \includegraphics[width=0.292\textwidth]{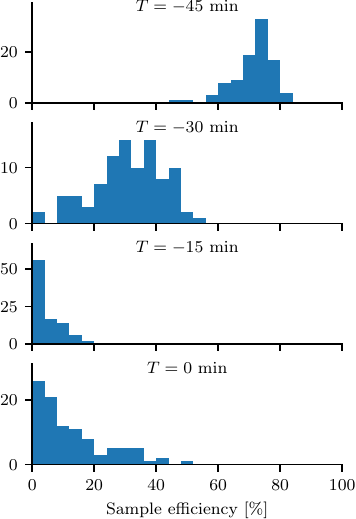}}
    \caption{
    Sample efficiencies for the injection studies. 
    (a) GW170817-like injections using different detector noise levels. (b) Injections using LVK design sensitivity PSDs. (c) Injections using CE PSDs.
    }
    \label{fig:sample-efficiencies}
\end{figure*}

We report sample efficiencies for all injections studies in Fig.~\ref{fig:sample-efficiencies}. Importance-sampled \textsc{Dingo-BNS} results are accurate even with low efficiency, provided that a sufficient \emph{absolute} number of effective samples can be generated. The efficiency nevertheless is a valuable diagnostic to asses the performance of the trained inference networks.

In LVK experiments, we find consistently high efficiencies, comparable to or higher than those reported for binary black holes~\cite{Dax:2022pxd}. As a general trend, we observe that higher noise levels (Fig.~\ref{fig:efficiency-GW170817-like}) and earlier pre-merger times (Fig.~\ref{fig:efficiency-DS} lead to higher efficiencies. This is because low SNR events generally have broader posteriors, which are simpler to model for \textsc{Dingo-BNS} density estimators. Furthermore, the GW signal morphology is most complicated around the merger, making pre-merger inference a much simpler than inference based on the full signal. 

For CE injections with GW170817-like parameters (Fig.~\ref{fig:efficiency-CE}), \textsc{Dingo-BNS} achieves extremely high efficiency for early pre-merger analyses but the performance decreases substantially for later analysis times. This effect can again be attributed to the increase in SNR, which is of $O(10^3)$ 15 minutes before the merger. Improving \textsc{Dingo-BNS} for such high SNR events will likely require improved density estimators~\cite{Wildberger2023flow} that can better deal with tighter posteriors. When limiting the SNR by increasing the distance prior (Tab.~\ref{tab:priors}), we find good sample efficiencies for an after-merger CE analysis that uses the full 4096~second long signal (Fig.~\ref{fig:efficiency-CE}). 

\subsection{Inference times}
The computational cost of inference with \textsc{Dingo-BNS} is dominated by (1) neural network forward passes to sample from the approximate posterior $\theta\sim q(\theta|d_{\Mcp},\Mcp)$ and by (2) likelihood evaluations $p(\theta|d)$ used for importance sampling. For 50,000 samples on an H100 GPU, (1) takes $\sim 0.370~\text{seconds}$ and (2) takes $\sim 0.190~\text{seconds}$, resulting in an inference time of less than $0.6~\text{seconds}$. The speed of the likelihood evaluations is enabled by using JAX waveform and likelihood implementations~\cite{Edwards:2023sak,Wong:2023lgb,Wouters:2024oxj}, combined with the heterodyning and multibanding step that we also use to compress the data for the \textsc{Dingo-BNS} network. We extend the open source implementations~\cite{Edwards:2023sak,Wouters:2024oxj} by combining NRTidalv1~\cite{Dietrich:2017aum,Dietrich:2018uni} with IMRPhenomPv2~\cite{Hannam:2013oca,Khan:2015jqa,Bohe:2016} as well as re-implementing the \textsc{Dingo} likelihood functions in JAX. We can jit the likelihood ahead of time since we evaluate a fixed number of waveforms at a fixed number of frequency bins. Thus we leave the jitting time (18 seconds) out of the timing estimate for importance sampling. This is in contrast to previous JAX-based GW works \cite{Wong:2023lgb,Wouters:2024oxj} which use a fiducial waveform (determined at inference time via likelihood maximization) to perform heterodyning. Likelihood evaluations can also be done without JAX, which takes less than 10 seconds on a single node with 64 CPUs for 50,000 samples. For the vast majority of \textsc{Dingo-BNS} analyses in this study, the sample efficiency is sufficiently high such that 50,000 samples correspond to several thousands of effective samples after importance sampling, enabling full importance sampling inference in less than a second. Note that these numbers refer to inference times, assuming data have already been provided to \textsc{Dingo-BNS}. Accelerating other aspects of LVK low-latency pipelines is critical for minimizing alert times~\cite{Chaudhary:2023vec}.

\subsection{PSD tuning}
Although most of the networks used in this study are trained with only a single PSD per detector, in practice we would generally train \textsc{Dingo-BNS} with an entire distribution of PSDs to enable instant tuning to drifting detector noise~\cite{Dax:2021tsq}. (This is not relevant to tests involving, e.g., design sensitivity noise.) Conditioning on the PSD makes the inference task more complicated, and therefore leads to slightly reduced performance. For example, when repeating the first injection experiment (Fig.~\ref{fig:efficiency-GW170817-like}) with a \textsc{Dingo-BNS} network trained with a distribution covering the entire second LVK observing run (O2), the mean efficiency is reduced from 57\% to 25\%. Such networks can in principle also be trained before the start of an observing run, by training with a synthetic dataset designed to reflect the expected noise PSDs~\cite{Wildberger:2022agw}.

\end{document}